\documentclass[aps,twocolumn,floats,prd,nofootinbib,10pt,longbibliography,superscriptaddress]{revtex4-1}
\usepackage{empheq}

\usepackage{comment}
\usepackage[dvips]{graphicx} %
\usepackage{graphicx,amsmath,amsfonts,amssymb,slashed,float,hyperref,empheq}
\usepackage[normalem]{ulem}
\usepackage{bbold,wasysym}
\usepackage{graphicx}
\usepackage{array,multirow}
\usepackage[utf8]{inputenc}
\usepackage{scalerel}

\usepackage[usenames,dvipsnames]{xcolor} 

\usepackage{soul}
\usepackage{bm}

\definecolor{RoyalBlue}{rgb}{0.25,.41,.88}
\definecolor{celestialblue}{rgb}{0.29, 0.59, 0.82}

\setstcolor{Blue}

\newcommand{\be}{\begin{equation}}
\newcommand{\ee}{\end{equation}}
\newcommand{\bea}{\begin{eqnarray}}
\newcommand{\eea}{\end{eqnarray}}
\newcommand{\Beq}{\begin{equation}\begin{aligned}}
\newcommand{\Eeq}{\end{aligned}\end{equation}}

\usepackage{color}
\usepackage{ifthen}
\newboolean{editorial}
\setboolean{editorial}{true}
\newcommand{\editorial}[2]{\ifthenelse{\boolean{editorial}}{\textcolor{red}{[\textsf{\textbf{{#1}}}: }\textcolor{blue}{\textsf{{#2}}}\textcolor{red}{]}}{}}

\usepackage{xcolor}

\begin{document}

\title{Primordial Black Holes from Kinetic Preheating}

\author{Peter Adshead}
\affiliation{Illinois Center for Advanced Studies of the Universe \& Department of Physics, University of Illinois at Urbana-Champaign, Urbana, IL 61801, U.S.A.}
\affiliation{Department of Physics and Astronomy,
University of Pennsylvania, 209 South 33rd St, Philadelphia, PA 19104}

\author{Eve Currens}
\affiliation{Department of Physics, Kenyon College, 201 N College Rd, Gambier, OH 43022, USA}

\author{John~T.~Giblin~Jr.}
\affiliation{Department of Physics, Kenyon College, 201 N College Rd, Gambier, OH 43022, USA}
\affiliation{Department of Physics/CERCA/Institute for the Science of Origins, Case Western Reserve University, Cleveland, OH 44106-7079, USA}
\affiliation{Center for Cosmology and AstroParticle Physics (CCAPP) and Department of Physics, Ohio State University, Columbus, OH 43210, USA}

\begin{abstract}
We demonstrate that violent kinetic preheating following inflation can lead to the formation of black holes in the early Universe. In $\alpha$-attractor models with derivative inflaton couplings, nonlinear amplification of field fluctuations drives large spacetime curvature and gravitational collapse shortly after inflation ends. Using fully general-relativistic lattice simulations, we find that these dynamics produce black holes with masses of order tens of grams at sub-horizon scales, without requiring large primordial curvature perturbations. Although such micro-black holes evaporate rapidly via Hawking radiation, their formation modifies the post-inflationary equation of state and their evaporation can successfully reheat the Universe before Big Bang nucleosynthesis. These results identify kinetic preheating as a new, efficient channel for black-hole production.

\end{abstract}

\maketitle
\section{Introduction}

The inflationary epoch not only explains the observed large-scale homogeneity of the Universe \cite{Guth:1980zm,Albrecht:1982wi, Linde:1983gd}, but also provides a natural origin for primordial fluctuations that seed cosmic structure \cite{Guth:1982ec, Hawking:1982cz, Bardeen:1983qw}. Yet, the transition from inflation to the hot Big Bang, reheating, remains one of the least understood phases in the early Universe. In many well-motivated models, the inflaton’s couplings to matter fields can drive non-perturbative and highly non-linear energy transfer, a process known as preheating \cite{Traschen:1990sw, Shtanov:1994ce, Kofman:1994rk, Kofman:1997yn, Greene:1997fu}.  Preheating generically leads to the production of a  high-frequency (MHz-GHz) stochastic gravitational wave background due to the generation of large gradients in the energy density \cite{Khlebnikov:1997di, Easther:2006gt, Easther:2006vd,Garcia-Bellido:2007nns, Garcia-Bellido:2007fiu, Dufaux:2007pt, Easther:2007vj, Bethke:2013vca, Dufaux:2010cf, Garcia-Bellido:2007nns,Figueroa:2016ojl, Figueroa:2017vfa}. When the transfer proceeds through derivative, or kinetic couplings rather than potential-like interactions, the resulting dynamics can be especially violent \cite{Adshead:2015pva, Cuissa:2018oiw, Adshead:2023nhk}, sourcing large gradients \cite{Adshead:2016iae}, strong gravitational waves \cite{Adshead:2018doq, Adshead:2019igv, Adshead:2019lbr,Weiner:2020sxn, Adshead:2023mvt, Adshead:2024ykw}, and as we demonstrate in this letter, gravitational collapse to black holes.

In previous work \cite{Adshead:2023nhk, Adshead:2024ykw}, we showed that kinetic preheating, which arises naturally in the conformal symmetry-based constructions of multifield $\alpha$-attractor inflationary models, can lead to rapid fragmentation of the inflaton condensate and the formation of localized, high-density regions. Using lattice simulations we previously showed that these configurations are highly inhomogeneous, producing stochastic gravitational-wave backgrounds with energy densities large enough to perturb the late-time expansion rate and lead to observational effects through shifts in the effective number of relativistic species, $N_{\rm eff}$.

In the present work, we extend this analysis to explore the ultimate nonlinear gravitational outcome of this process. We find that, under generic conditions, the violent inhomogeneities generated during kinetic preheating can seed gravitational collapse and form black holes with masses of order tens of grams. These micro-black holes emerge dynamically from the field fluctuations themselves, without requiring any special initial conditions, and constitute a new and robust pathway to black-hole production in the early Universe.

Unlike primordial black holes (PBHs) \cite{Hawking:1971ei, Carr:1974nx, Carr:1975qj, Green:1997sz, Musco:2004ak, Khlopov:2008qy, Musco:2012au, Harada:2013epa} formed from large curvature perturbations during inflation \cite{Garcia-Bellido:2017mdw, Byrnes:2018txb, Sasaki:2018dmp, Bhattacharya:2019bvk,Martin:2019nuw}, the black holes produced here originate from the intrinsically nonlinear, post-inflationary dynamics of preheating. Their formation reflects the strong coupling between field gradients and spacetime curvature in the fully relativistic regime, which we capture using GABERel \cite{Giblin:2019nuv,Adshead:2023mvt}  --- an extension of GABE \cite{Child:2012qg} that evolves the metric and matter fields self-consistently in the Baumgarte-Shapiro-Shibata-Nakamura (BSSN) scheme of numerical relativity \cite{Baumgarte:1998te, Shibata:1995we}. The resulting gravitational collapse occurs at sub-horizon scales and within a few oscillations of the inflaton field, highlighting the extreme efficiency of energy localization during kinetic preheating.

The black holes produced in this scenario are far too light to survive to the present day, evaporating via Hawking radiation shortly after formation. Nevertheless, their transient existence can have significant cosmological implications. Their evaporation is sufficient to reheat the Universe without any further couplings between the inflationary sector and the standard model sector. Further, a PBH dominated phase may lead to nonthermal particle production, modify the post-inflationary equation of state, or imprint characteristic features in the stochastic gravitational-wave spectrum.  More broadly, these results establish kinetic preheating as a qualitatively new channel for black-hole formation in the early Universe.


\section{The Model\label{sec:model}}

We consider a kinetic-preheating scenario for an axion-diliton inflationary model
\begin{align}\label{eq:lagrangian}
\mathcal{L}= -\frac{M_{\rm Pl}^2}{2}R -\frac{1}{2}\left(\partial\varphi\right)^{2} - \frac{W(\varphi)}{2}(\partial\chi)^2 - V(\varphi),
\end{align}
in which the dilaton, $\varphi$, is kinetically coupled to the axion, $\chi$, through an exponential dilaton-like kinetic coupling, $W(\varphi) = e^{2\varphi/\mu}$.

For this analysis, we consider the the asymmetric E-model $\alpha$-attractor potential \cite{Kallosh:2013maa,Linde:2018hmx}
\begin{equation} \label{eq:Emodel}
V = \frac{m^2\mu^2}{2}\left(1-e^{-\frac{\varphi}{\mu}}\right)^2.
\end{equation}

For small values of $\mu$, preheating is extremely efficient and, as demonstrated in Friedmann-Lem{a}\^{i}tre-Robertson-Walker (FLRW) spacetime in refs.\ \cite{Adshead:2024ykw, Adshead:2023nhk}, leads to large spikes in the density contrast at several scales and a loud background of stochastic gravitational waves at high frequencies (GHz). In this letter, we extend the results of these works to include the effects of nonlinear gravitation to allow these overdense regions to undergo gravitational collapse.

To implement fully nonlinear gravity we apply the BSSN formalism \cite{Shibata:1995we, Baumgarte:1998te} (see, e.g. \cite{Baumgarte:2010ndz}) which employs a 3+1 decomposition of spacetime, 
\begin{align}
    ds^2 = \left(-\alpha^2+\beta_i\beta^i\right){\rm d}t^2 + 2\beta_i{\rm d}t{\rm d}x^i + e^{4\phi}\bar{\gamma}_{ij}{\rm d}x^i{\rm d}x^j,
    \label{eqn:bssn-metric}
\end{align}
where the lapse, $\alpha$, and shift, $\beta$ are pure gauge degrees of freedom whose dynamical equations are chosen to keep the 3-dimensional surfaces of the simulation purely spatial.  The evolution of the metric is governed by the {\sl extrinsic curvature}, $K_{ij} = \tilde{A}_{ij} - \delta_{ij}K/3$. In a pure homogeneous and isotropic Friedman-Lema\^itre-Robertson-Walker (FLRW) universe, the mean curvature is related to the Hubble parameter via $K = -3H$.  The full set of non-linear differential equations that define the evolution of the metric degrees of freedom can be found, e.g. in \cite{Baumgarte:2010ndz}.

The kinetic coupling modifies the standard equations of motion for scalar fields.  We define
\begin{equation}
    \Pi \equiv \frac{1}{\alpha}\left(\partial_t\phi-\beta^i\partial_i\varphi\right),
\quad 
    \Theta\equiv \frac{1}{\alpha}\left(\partial_t\chi-\beta^i\partial_i\chi\right),
\end{equation}
which can be used to write the equations of motion for the two fields,
\begin{align}
    \nonumber
    \partial_0 \Pi =  &\beta^k \partial_k \Pi + \alpha K\Pi + \gamma^{ij} \partial_i \alpha \partial_j \varphi - \alpha \gamma^{ij} \Gamma^k_{ij} \partial_k \varphi \\
    &-\frac{\alpha}{2} \frac{\partial W}{\partial \varphi} \left(\gamma^{ij}\partial_{i}\chi \partial_{j}\chi - \Theta^2\right) -\alpha\frac{\partial V}{\partial\varphi},\label{eq:eomphi}
    \\
    \nonumber
    \partial_0 \Theta=&\beta^k\partial_k\Theta+\alpha K \Theta +\gamma^{ij}\partial_i\alpha\partial_j\chi - \alpha\gamma^{ij}\Gamma^k_{ij}\partial_k\chi \\
    &+ \frac{\alpha}{W}\frac{\partial W}{\partial \varphi}\left(\gamma^{ij}\partial_i\varphi\partial_j\chi - \Pi\Theta\right).
    \label{eq:eomchi}
\end{align}
In practice, the numerical system is more stable if the gradients of the fields evolve independently.  For example, we define $\psi_i \equiv \partial_i\varphi$ and evolve
\begin{equation}
    \partial_0\psi_i = \beta^j\partial_j\psi_i+\psi_j\partial_i\beta^j-\alpha\partial_i\Pi - \Pi\partial_i\alpha\,,
\end{equation}
while substituting $\psi_i$ into eqs.~\ref{eq:eomphi} and \ref{eq:eomchi}.

We find the homogeneous values of $\varphi_0$ and $\dot{\varphi}_0$ by numerically integrating the homogeneous Klein-Gordon equations during inflation for a pure FLRW universe, using the same method as refs.\ \cite{Adshead:2023nhk,Adshead:2024ykw}.

The inhomogeneous initial conditions are set as in ref.\ \cite{Giblin:2019nuv}, where the two-point correlation function of each scalar field is given Bunch-Davies fluctuations,
\begin{equation}
\langle\left|\delta\varphi_k\right|^2\rangle = \frac{1}{2a^2\omega_k},
\quad 
    \langle\left|\delta\chi_k\right|^2\rangle = \frac{1}{2W(\varphi)a^2\omega_k},
\end{equation}
where $\langle\left|f_k\right|^2\rangle$ is the ensemble average. 
We also apply a window function (as in \cite{Giblin:2019nuv}) to reduce the power in high-frequency modes that we expect to be outside of the tachyonic instability.  The fluctuations of the gravitational sector are set mode-by-mode using perturbation theory. We first calculate $\delta \rho_k$ and $\delta^{ij}\partial_i T_{0j}$ on the initial slice then solve the associated Poisson equations \cite{Giblin:2019nuv} for the Bardeen potential $\Phi$.  The full set of initial conditions for the gravitational fields are $\alpha = 1 + \Phi$,$\phi = -\Phi/2$, and
\begin{align}
    K &= -3H + 3\left(\dot{\Phi} + H\Phi\right),
\end{align}
alongside the choices of $\beta^i=\tilde{A}_{ij} = 0$, $\bar{\gamma}_{ij} =\delta_{ij}$.  We employ a Bona-Mass\'o slicing condition to the lapse, $\partial_t \alpha = -2\alpha (K - \langle K\rangle)$, and use the standard hyberbolic gamma driver slicing condition on the shift, $\partial_t \beta_i = 3B^i/4+\beta^j \partial_j\beta^i$ and $\partial_t B = \partial_t \Gamma^i  - \eta B^i/2 + \beta^j\partial_j B^i$, with $\eta = 10^2$ \cite{Baumgarte:2010ndz}. For configuration-space quantities, $\langle \cdots \rangle$ is a spatial average over constant-$t$ hypersurfaces.

\section{Results}\label{sec:results}

To demonstrate the formation of black holes, we focus on a single value of $\mu$ chosen among those tested in \cite{Adshead:2023nhk, Adshead:2024ykw}; for $\mu\approx4.68\times10^{-2} M_{\rm pl}$, which also implies $m = 8.04\times10^{-6}\,M_{\rm pl}$, we anticipate efficient preheating.  The results we present here are from a simulation that begins one-half an e-folding before the end of inflation, for consistency with \cite{Adshead:2023mvt,Adshead:2024ykw}; in that work we chose the box to be Hubble-scale at the end of inflationto resolve the tachyonic instability.  Here, we choose a smaller box to resolve the high-frequency modes while still being able to see the tachyonic instability.  To accomplish this, we take $L= H_{\rm end}^{-1} e^{-0.5}/5\approx 6.72 \, m^{-1}$, which corresponds to a box size of $H_{\rm end}^{-1}/5$ as the end of inflation. The grid is taken to have $N^3=256^3$ points.  With a timestep of $\Delta t = L/N/30$. We express time in units of the Hubble scale at the beginning of the simulation, $H_*^{-1} \approx 0.019\,m^{-1}$, which represents the fundamental time-scale of the problem. 

To test for convergence, we ran lower-resolution simulations ($N^3=128^3$) as well as simulations with various box-sizes, ranging from $L\approx6.72\,m^{-1}$ to $L\approx33.6\, m^{-1}$, as well as several slicing conditions, including those designed to better resolve horizons \cite{Staley:2011ss}. We also performed a simulation in which we reduce the timestep by a factor of ten (to $\Delta t = L/N/300$) after $t\approx 1.70\,H_*^{-1}$ to ensure convergence during the collapsing stage.  In all of these simulations we found the same behavior; we choose this simulation since it showed the longest numerical stability with sufficient time-resolution to see the relevant processes of preheating and the formation of the black hole. 

\begin{figure}
\includegraphics[width=\columnwidth]{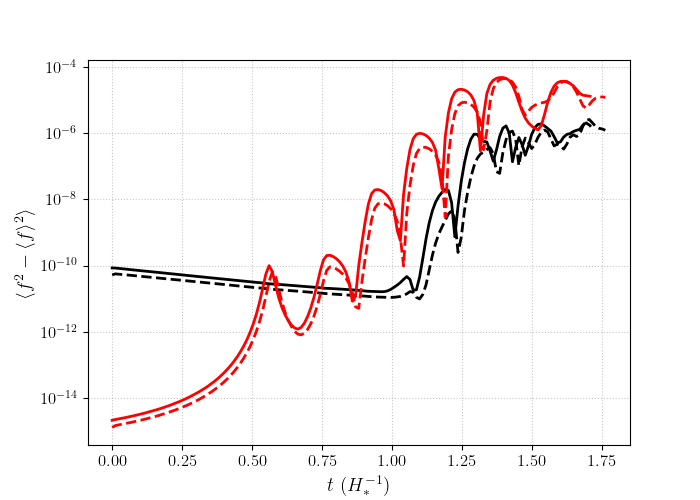}
    \caption{\label{fig:meansvariances}The variances of the inflation (black), $\varphi$, and the axion (red), $\chi$ over time for the run presented here (solid) and a corresponding FLRW simulation (dashed).  The rise of the variance of the axion represents the phase of kinetic preheating which extends from $\approx 0.5 \,H_*^{-1}$ until $t\approx 1.25\,H_*^{-1}$.}
\end{figure}

\begin{figure}
    \includegraphics[width=\columnwidth]{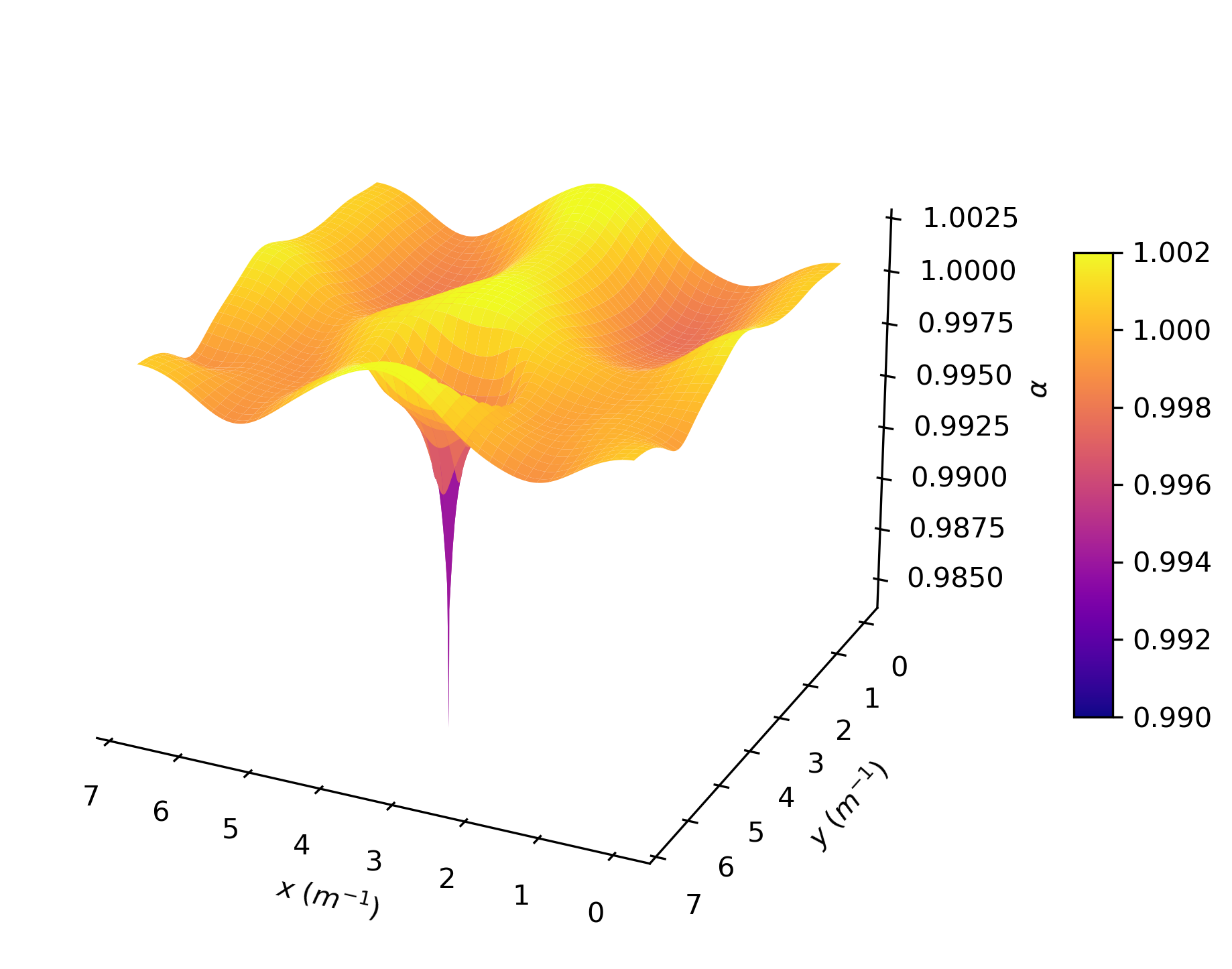}
    \caption{\label{fig:funnel} The lapse, $\alpha$, at $t\approx 1.7050 H_*^{-1}$ right before gravitational collapse begins.}
\end{figure}

To validate the fully nonlinear simulation, we  calculate the statistics of the $\phi$ and $\chi$ fields in Fig.~\ref{fig:meansvariances}.
The variances of the field show consistency with the FLRW simulation throughout the time when the field is in the linear regime as well as during the tachyonic reheating phase and during the phase of nonlinear evolution, demonstrating that the field equations are consistent with previous studies.  
\begin{figure*}
    \includegraphics[width=\textwidth]{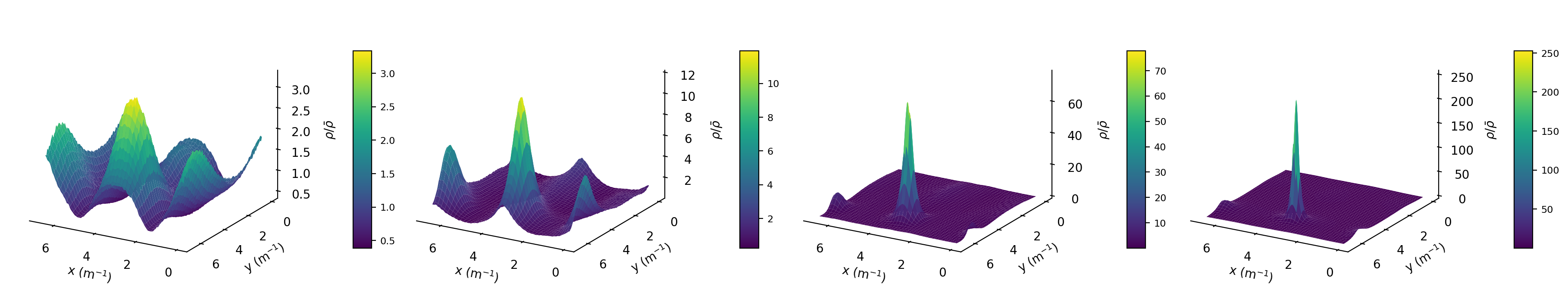}
    \caption{\label{fig:evolutionofrho}Two-dimensional slices of the density, $\rho/\langle \rho\rangle$ over several slices from the end of the tachyonic resonance period until the gravitational collapse begins.  From left to right, these are at $t\approx 1.3730 \,H_*^{-1},1.5332 \,H_*^{-1},1.5904\, H_*^{-1},$ and $ 1.6705 \,H_*^{-1}$. Note the increasing scale of the vertical axis over these four slices.}
\end{figure*}Fig.~\ref{fig:evolutionofrho} shows how the density, $\rho/\langle\rho\rangle$ evolves as the tachyonic instability creates large density contrasts.  Fig.~\ref{fig:funnel} demonstrates how the lapse, $\alpha$, begins to diverge from unity as the density contrast grows, just before it gravitationally collapses.

\begin{figure*}
    \includegraphics[width=\textwidth]{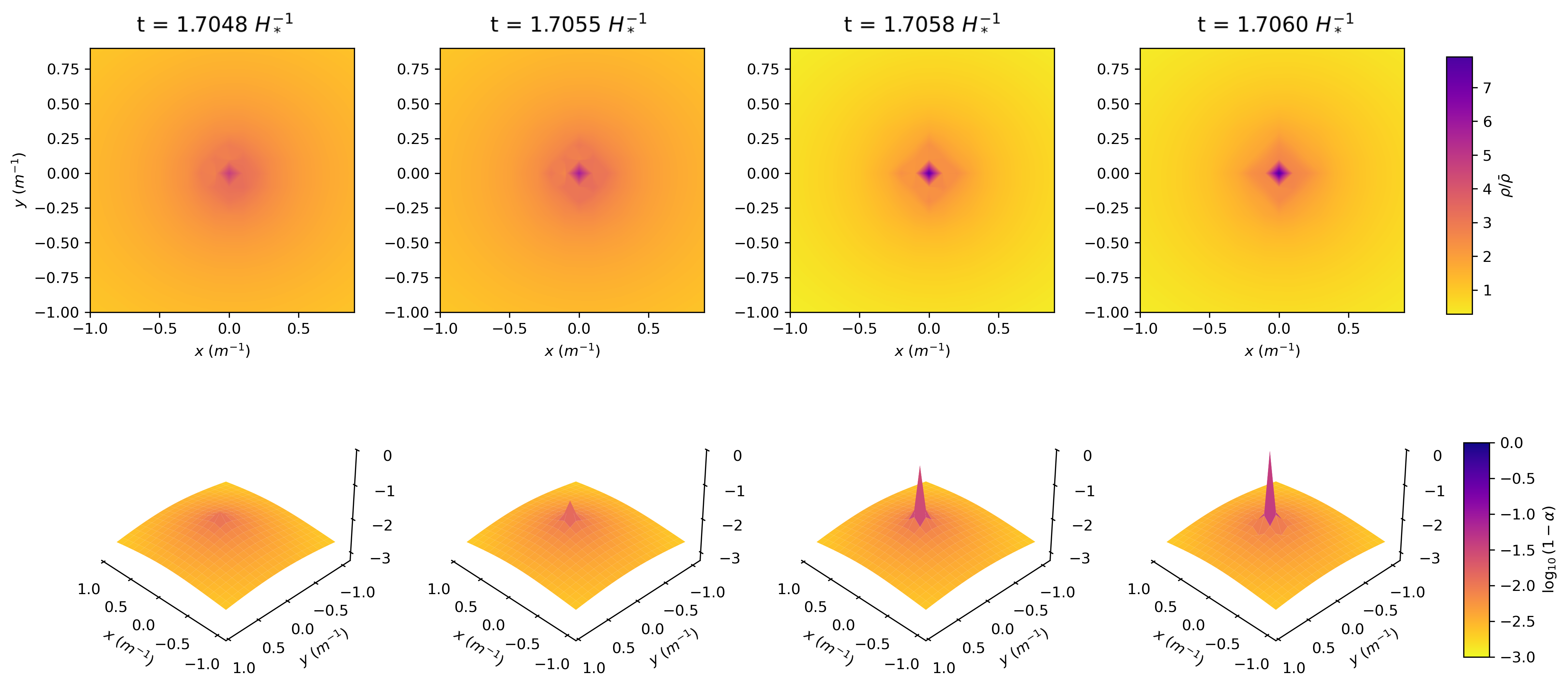}
    \caption{\label{fig:collapse}The density contrast (top panels) and the scaled lapse, $\log_{10}(1-\alpha)$, (bottom panels) during the time when the over-dense region is collapsing. We scale the lapse to emphasize that it is departing significantly from one and we plot only the region surrounding the emerging black hole.}
\end{figure*}

As the simulation continues to evolve, the overdense regions begins to gravitationally collapse, as can be seen in the panels in Fig.~\ref{fig:collapse}. 
The time scale over which this collapse occurs is very short compared to the timescale of the preheating instability. Within $\Delta t \approx 0.002 H_*^{-1} $ we see that the density contrast grows by several orders of magnitude while the value of the shift goes to zero.  

\begin{figure}
    \includegraphics[width=\columnwidth]{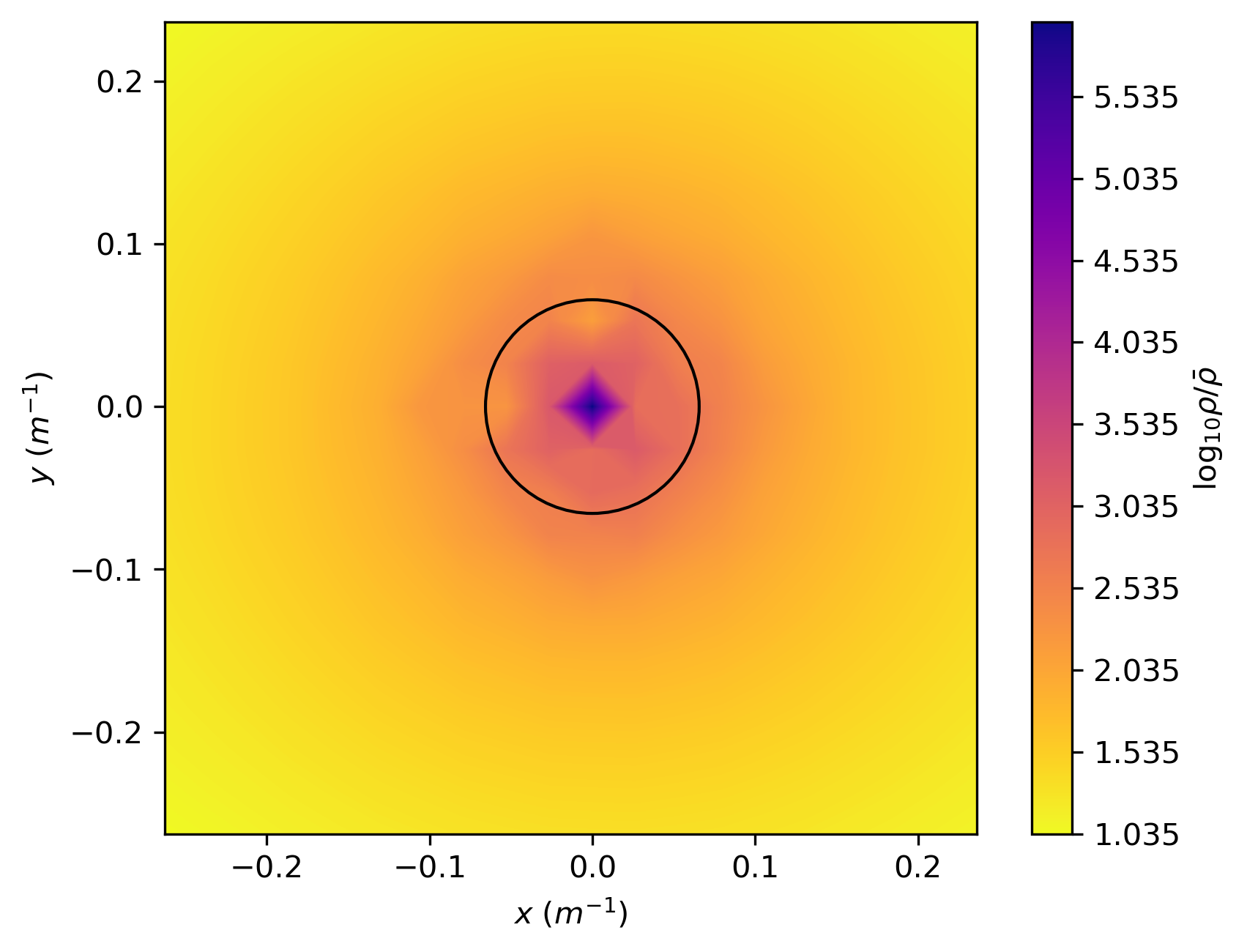}
    \caption{\label{fig:horizon}The density contrast at the time of the horizon formation, $t\approx 1.7063 \,H_*^{-1}$. The black circle indicates the apparent horizon.}
\end{figure}

At $t\approx 1.7063 H_*^{-1}$ we see the formation of a black hole horizon.  We identify apparent horizons by searching for trapped surfaces \cite{Thornburg:2003sf}; we evaluate the expansion
\begin{equation}
\label{eq:expansion}
    \Theta \equiv \nabla_i n^i + K_{ij} n^i n^j - K \,,
\end{equation}
on spherical surfaces surrounding the overdense region.  The outermost surface on which the expansion vanishes is the apparent horizon, $r_\star$. While the size collapsed region is small compared to our box, there are still many points inside $r_\star$.  For our case, the coordinate apparent horizon is located at $r_\star\approx 2.5\Delta x \approx 0.65 m^{-1}$ as seen in Fig.~\ref{fig:horizon}.
Finally, we estimate the mass of the black hole by calculating the density in the collapsing region.  We look at several successive slices from $t\approx 1.66\,H_*^{-1}$ until $t\approx 1.70\,H_*^{-1}$ and sum over the region where $(\rho-\langle\rho\rangle)/\langle\rho\rangle > 25$.  For this simulation, $M_{\rm BH} = 30\,{\rm g}$, which is approximately $3.5$\% of the total mass of the simulation at that time.

\section{Reheating from Primordial Black-Hole Evaporation}\label{sec:reheating}

Once PBHs are formed, the subsequent cosmological evolution depends on their evaporation history.  PBHs behave as nonrelativistic matter, so if their initial energy fraction at the time of formation $t_f$, $\beta_f \equiv \rho_{\rm PBH}(t_f)/\rho_{\rm tot}(t_f)$ is not exponentially small, they quickly come to dominate the energy density. Our simulations indicate that $\beta_f\lesssim 0.05$. Assuming a radiation equation of state during preheating, the PBH fraction grows as $\Omega_{\rm PBH}\propto a$, and PBH-reheating equality occurs at
\begin{align}
t_{\rm eq}\simeq \frac{t_f}{\beta_f^2}.
\end{align}
Subsequently a matter-dominated phase follows until the black holes evaporate via Hawking radiation \cite{Hawking:1975vcx, Page:1976df, Anantua:2008am, Zagorac:2019ekv} at
\begin{align}
t_{\rm evap} =\frac{10240\pi}{g\, m_{\rm pl}^4}M_{\rm BH}^3\simeq 1.7\times10^{-27}{\rm s}
\left(\frac{M}{\rm g}\right)^3
\left(\frac{100}{g_\star}\right),
\end{align}
where $g$ is the greybody-weighted effective number of degrees of freedom --- $g = 15.25$ for the full standard model.

For the $\mathcal{O}(10-100)\,{\rm g}$ black holes produced during kinetic preheating, $t_{\rm evap}\sim10^{-24}-10^{-21}\,{\rm s}$, corresponding to an extremely short but genuine PBH-dominated epoch lasting $\Delta N_{\rm MD}\simeq (2/3)\ln(t_{\rm evap}/t_{\rm eq})\sim10-20$ e-folds.

The evaporation of these light PBHs rapidly converts their mass into a relativistic plasma, reheating the Universe to
\begin{align}
T_{\rm reh}^{\rm (PBH)} \simeq
0.87~{\rm MeV}
\left(\frac{g_\star}{100}\right)^{-1/4}
\left(\frac{10^9{\rm g}}{M}\right)^{3/2},
\end{align}
so that $M_{\rm BH}\simeq30\,{\rm g}$ yields $T_{\rm reh}\sim10^{8}\,{\rm GeV}$.
Because $t_{\rm evap}\ll1\,{\rm s}$, this reheating occurs well before Big-Bang nucleosynthesis, leaving standard light-element abundances unaffected.  Hawking emission distributes roughly $f_g\sim10^{-2}$ \cite{Page:1976df} of the PBH energy into gravitons, giving a negligible $\Delta N_{\rm eff}^{\rm (GW)} \lesssim 10^{-3}$. However,  any additional decoupled light species attains approximately the same temperature as the visible Standard Model, and therefore contributes  $\Delta N_{\rm eff} \simeq 0.027 (g^b_X +7 g^f_X/8)$, where $g^{b,f}_X$ counts the internal degrees of freedom for bosons or (Weyl) fermions, respectively.  In particular, an additional three right-handed (sterile) neutrinos added to the Standard Model contribute  $\Delta N_{\rm eff}\approx 0.14$. While this is within current bounds, it is well within the range targeted by upcoming experiments \cite{CMB-S4:2016ple,Abazajian:2019eic,CMB-S4:2022ght}.

Thus, evaporation of the micro-black holes formed during kinetic preheating provides a natural and efficient reheating mechanism: it restores a hot radiation bath at $T_{\rm reh} \sim 10^{8-10}{\rm GeV}$, precedes BBN by many orders of magnitude, and leaves only a minuscule residual contribution to relativistic energy density today.
Kinetic preheating after $\alpha$-attractor inflation therefore provides a realization of the scenarios discussed in \cite{RiajulHaque:2023cqe}.

\section{Discussion and Conclusion}\label{sec:conclusions}

In this letter, we have demonstrated that black holes can be formed during kinetic preheating after $\alpha$-attractor inflation.  Using numerical simulations we have shown that the kinetic coupling between a dilaton inflaton, and an axion-reheaton, leads to a violent tachyonic instability which sources large localized  density fluctuations that subsequently  undergo gravitational collapse to black holes.

Our simulations use a fixed-grid which limits our ability to resolve both the scale of the preheating instability and the scales associated with the black hole as the horizon forms.  While we have several points inside this region, a next step would be to further study this collapse using an adaptive mesh scheme, such as those used in \cite{Andrade:2021rbd}. We generically observe around 3.5\% of the energy density in our simulations collapsing into a single black hole, and anticipate at least one black hole per horizon volume at the end of inflation.

After collapse into black holes, the Universe evolves as radiation dominated until reheaton-PBH equality, where it subsequently evolves in a matter dominated phase that lasts 10-20 $e$-foldings. The Universe is then reheated as these black holes evaporate via Hawking radiation. Reheating through Hawking radiation populates all gravitationally coupled degrees of freedom without any direct couplings between the inflationary sector and the standard model.


\acknowledgments

We thank Josu Aurrekoetxea, Thomas Baumgarte, Katy Clough, Mary Gerhardinger, Amanda Miller, Eugene Lim, Avery Tishue and Chul-Moon Yoo for extremely helpful discussions and correspondence.  J.T.G. \ is  grateful for the hospitality of the Illinois Center for Advanced Studies of the Universe at the University of Illinois at which some of this work was conducted. P.A.\ is supported in part by the United States Department of Energy, DE-SC0015655. 
J.T.G. and E.C.\ are supported in part by the National Science Foundation, PHY-2309919.  

\bibliography{kin_preheat}
\end{document}